\documentclass[
 reprint,
 bibnotes,
 amsmath,amssymb,
 aps,prd
]{revtex4-1}

\usepackage{graphicx}
\usepackage{epstopdf}
\usepackage{dcolumn}
\usepackage{bm}
\usepackage{hyperref}

\begin{document}

\title{Modeling of Energy Distributions in Pseudo-Rest Frame Analyses of Two-Body Decays with Missing Particles}

\author{J. A. Colorado-Caicedo}
\email[Email address: ]{johan.colorado@cinvestav.mx}
\affiliation{Centro de Investigacion y de Estudios Avanzados del Instituto Politecnico Nacional, Av. IPN 2508, San Pedro Zacatenco, Mexico City, 07360}
\author{C. Lizama-Garcia}
\email[Email address: ]{carlos.lizama@cinvestav.mx}
\affiliation{Centro de Investigacion y de Estudios Avanzados del Instituto Politecnico Nacional, Av. IPN 2508, San Pedro Zacatenco, Mexico City, 07360}
\author{E. De La Cruz-Burelo}
\email[Email address: ]{e.delacruz.burelo@cinvestav.mx}
\affiliation{Centro de Investigacion y de Estudios Avanzados del Instituto Politecnico Nacional, Av. IPN 2508, San Pedro Zacatenco, Mexico City, 07360}

\begin{abstract}
In this study, we introduce a parametric function designed to describe the energy distribution of the observed particle within the framework of two-body decays involving one undetected particle, analyzed using the pseudo-rest frame approximation. While we illustrate its effectiveness through the specific case study of the Lepton Flavor Violating decay $\tau \rightarrow l+\alpha$, this parametric function is broadly applicable to a wide range of pseudo-rest frame method-related searches involving undetected particles. Remarkably, it requires only a single simulation to account for the smearing effects resulting from the pseudo-rest frame approximation.
The uniqueness of this function lies in its dependency on the mass of the undetected particle, enabling continuous exploration of the mass parameter space. We validate the performance of our parametric function using simulated datasets and find that it exhibits comparable performance to traditional simulation-based methods. Notably, our approach offers the distinct advantage of accommodating any mass value for the undetected particle without the need for multiple simulations.
\end{abstract}

\maketitle

\section{Introduction}

The quest for particles beyond the confines of the Standard Model (SM) remains vital in extending our understanding of nature. Numerous theoretical frameworks assert the existence of elusive particles, which either interact weakly with ordinary matter or evade direct detection altogether~\cite{Beacham:2019nyx, Lanfranchi:2020crw}. This pursuit has prompted a vibrant field of scientific inquiry, leading to novel techniques, particularly in measuring the mass of yet-to-be-detected particles involving decays with missing energy~\cite{Barr:2010zj, HarlandLang:2012gn, Christensen:2014yya, Xiang:2016jni, DeLaCruz-Burelo:2020ozf, Guadagnoli:2021fcj}.

Despite significant advancements, many existing methods introduce variables sensitive to the mass of the hidden particle but lack a parametric expression to accurately describe the changes in these variables as a function of the new particle mass. That is a precise characterization of the underlying probability density function (PDF) governing the behavior of these variables. This gap forces researchers to rely on computationally intensive simulations for the construction of these PDFs, an approach constrained by statistics and limited in exploring the potential mass values for the new particle.

An illustrative example is the search for a hypothetical invisible new light $\alpha$ boson, postulated to address the fermion mass hierarchy problem~\cite{Grinstein:1985rt}. As a feature of various SM extensions~\cite{Feng:1997tn, Heeck:2016xkh, Altmannshofer:2016brv, Asai:2018ocx, Ibarra:2021xyk}, the $\alpha$ boson is sought in the charged Lepton Flavor Violating (LFV) decay $\tau \rightarrow l+\alpha$. Early efforts by MARK-III~\cite{MARK-III:1985iqj} and ARGUS~\cite{Albrecht:1995ht} have paved the way for recent studies by Belle II~\cite{Tenchini:2020njf, Belle-II:2022heu}. 

In the mother particle rest frame, the energy of the lepton arising from a two-body decay, such as $\tau \rightarrow l+\alpha$, can be expressed as $E = (m_{\tau}^2 - m_{\alpha}^2 + m_{l}^2)/2m_{\tau}$, where $m_{\tau}$, $m_{\alpha}$, and $m_{l}$ denote the masses of the tau, the $\alpha$ boson, and the lepton, respectively. This relation provides a potential path for determining the mass of the undetected particle $\alpha$ by measuring the energy of the lepton in the tau rest frame. However, practical implementation is hindered by the unobservable decay products in tau decays, rendering the transformation to the tau rest frame unfeasible.
 
\begin{figure}[ht]
\begin{center}
\includegraphics[width=8cm]{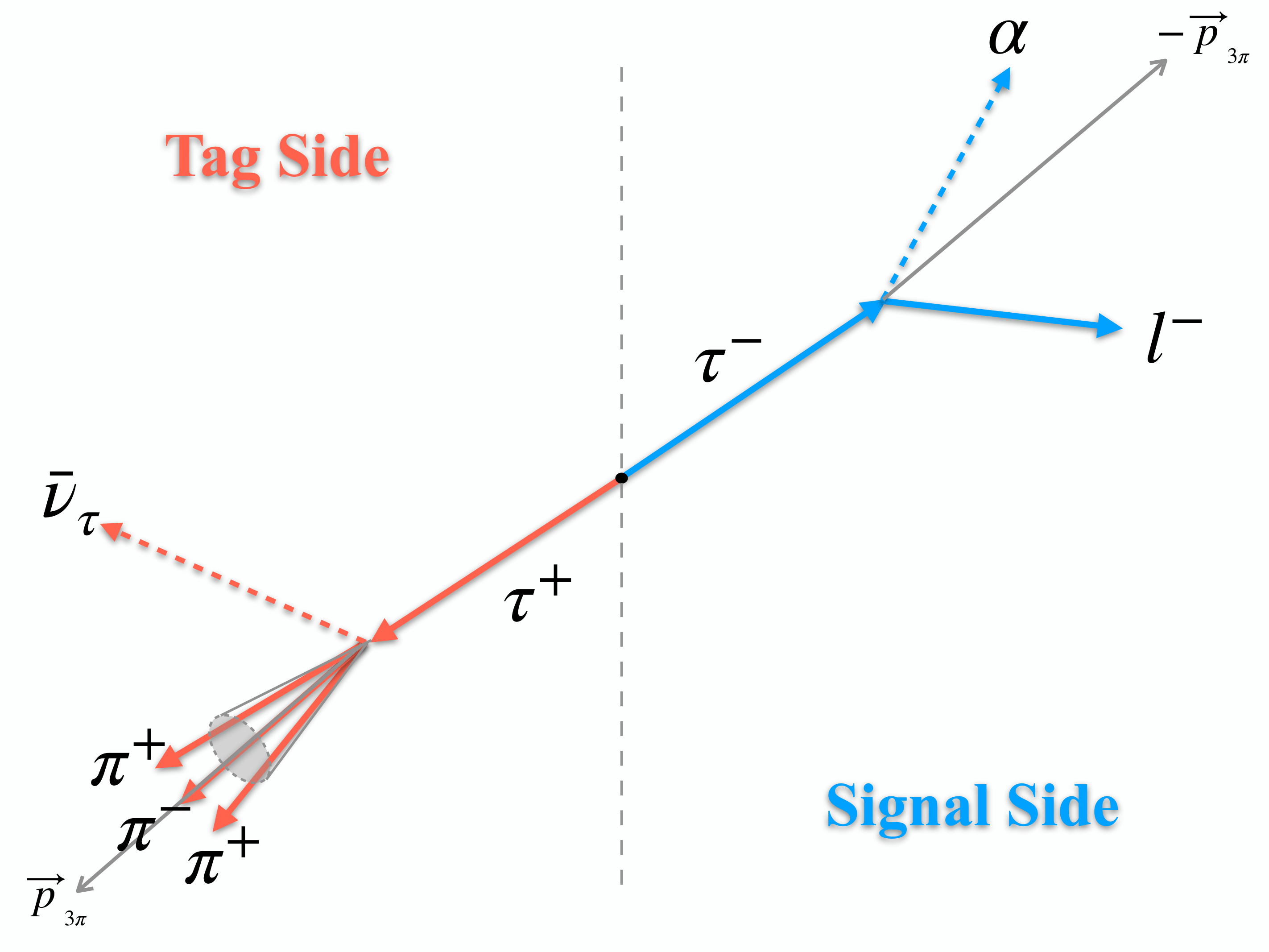}
\caption{Tau pair decay in a 3$\times$1 topology, with $\tau^{+} \to \pi^{+} \pi^{-} \pi^{+} \overline{\nu}_{\tau}$ decays on the tag side and $\tau^{-} \to \ell^{-}+\alpha$ on the signal side. Here \(l\) can be either an electron or a muon, and $\alpha$ is a particle that escapes detection.}\label{fig0}
\end{center}
\end{figure}

An approximation known as the pseudo-rest frame~\cite{Korolko:1994cz} has been employed by ARGUS and Belle II to address this limitation. In the center-of-mass frame (cms), the direction of the tau in $\tau \rightarrow l+\alpha$ decays is approximated by the reversed direction of the momentum of the observed particles on the decay of the other tau, as illustrated in Figure \ref{fig0}. Yet, this approach still falls short, as it lacks a comprehensive model to describe the lepton energy distribution in the pseudo-rest frame as a function of the $\alpha$ particle mass.
Due to this, both ARGUS and Belle II employed a simulation-based template method for their analyses, which specifically considered only a few discrete mass values: $0$, $0.5$, $0.7$, $1.0$, $1.2$, $1.4$, and $1.6$ GeV.

In the context of this paper, let us consider the specific example of $\tau \rightarrow l+\alpha$ decay. We present a straightforward procedure aimed at constructing a parametric function that characterizes the distribution of lepton energies within the pseudo-rest frame. This method, although conceptually simple, holds significant utility for modeling the energy distribution of any two-body decay occurring within a pseudo-rest frame, especially when such decay involves undetected particles, whether they belong to the SM or beyond.

A noteworthy improvement here is the parametric function, which facilitates a continuous search for the mass of the undetected particle, unburdened by the computational demands of conventional simulations. This approach liberates researchers from the constraint of discrete mass values, enabling a more comprehensive and nuanced exploration of the mass parameter space. We demonstrate the effectiveness of our method by comparing it with existing simulation-based approaches, revealing its equivalency to template-based methods. Notably, our approach offers the advantage of accommodating any mass value for the undetected particle without requiring more than one simulation run for the estimation of smearing effects arising from the pseudo-rest frame approximation.

The paper is organized as follows: Section \ref{sec:analyticalF} outlines the methodology to derive the parametric function for $\tau \rightarrow l+\alpha$ decays; Section \ref{sec:results} presents the results and contrasts them with existing simulation-based models; Section \ref{sec:conclusion} concludes with reflections on the broader impacts and implications of our research.

\section{Model for $\tau \rightarrow l+\alpha$ Decays}\label{sec:analyticalF}

\begin{figure}[!htb]
\centering
\includegraphics[width=8cm]{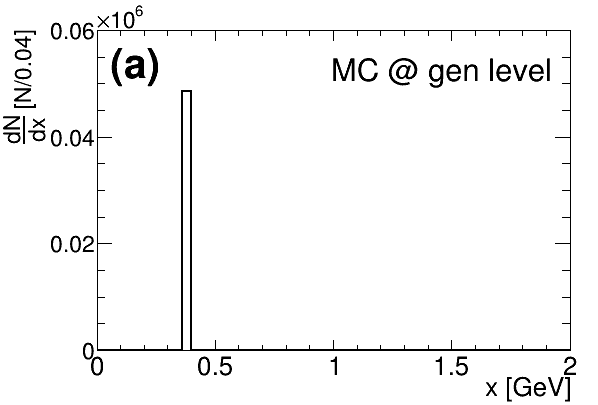}
\includegraphics[width=8cm]{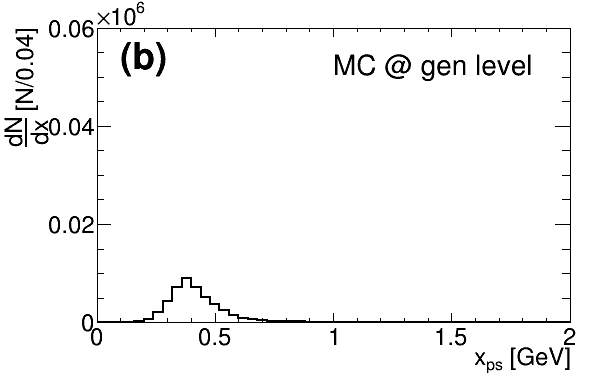}
\caption{Normalized lepton energy distribution for $\tau \to \ell+\alpha$ decays with $m_{\alpha}=1.4$ GeV in (a) the tau rest frame and (b) the tau pseudo-rest frame.}\label{fig3}
\end{figure}

To construct a parametric model for the lepton energy distribution in the tau pseudo-rest frame, we examine the $\tau^{+}\tau^{-}\to (\pi^{+}\pi^{-}\pi^{+}\overline{\nu}_{\tau})(l^{-}\alpha)$ decays searched by ARGUS and Belle II, as shown in Figure~\ref{fig0}. These decays involve pions on the tag side and a lepton on the signal side. The expected lepton distribution in the tau rest frame is expressed as
\begin{equation}\label{dNdx_2body}
 \frac{dN(x)}{dx} \sim \delta(x - b(m_{\alpha})),
\end{equation}
Here, $x=2E/m_{\tau}$ represents the normalized lepton energy in the tau rest frame, and
\begin{equation}\label{xrf_fix}
 b(m_{\alpha}) = \frac{m_{\tau}^2 - m_{\alpha}^2 + m_{l}^2}{m_{\tau}^2}.
\end{equation}
Figure~\ref{fig3} illustrates the $dN/dx$ distribution\footnote{Throughout the paper, only the decays $\tau \to e+\alpha$ are considered as an example.} in both the (a) tau rest frame and the (b) tau pseudo-rest frame. The comparison in Fig. \ref{fig3} highlights the smearing in the $b(m_{\alpha})$ value due to the transformation to the pseudo-rest frame. 

To analyze the smearing effect introduced by the use of the pseudo-rest frame, let us consider the Lorentz transformations of the lepton energy from the cms frame, $E_{cms}$, to the tau rest frame ($E$), and to the pseudo-rest frame ($E_{ps}$):
\begin{equation}\label{LTEnergy}
 E = \gamma(E_{cms} - \beta P_{cms}
 \cos\theta),  
\end{equation}
\begin{equation}\label{LTEnergyPS}
  E_{ps} = \gamma(E_{cms} - \beta P_{cms}
 \cos\theta_{ps}),    
\end{equation}
where $\beta$
is the velocity of the moving frame. Here, $P_{cms}$ represents the lepton momentum in the cms frame, and $\theta$ is the angle between the boost direction and the lepton momentum. Furthermore, $\gamma=(1-\beta^2)^{-1/2}$.

With $x_{ps}=2E_{ps}/m_{\tau}$, from Eq.~(\ref{LTEnergy}) and Eq. (\ref{LTEnergyPS}), we obtain:
\begin{equation}\label{DeltaX}
 x_{ps} - x = \frac{2P_{cms}}{m_{\tau}}\gamma\beta \Delta\cos\theta,
\end{equation}
where $\Delta\cos\theta=\cos\theta - \cos\theta_{ps}$. Considering $E_{cms} \gg m_{l}$, $2P_{cms}/m_{\tau} \approx 2E_{cms}/m_{\tau}=x_{cms}$, we have:
\begin{equation}\label{xps}
    x_{ps}\approx x + x_{cms}\gamma\beta \Delta\cos\theta
\end{equation}
Similarly, with $x\approx 2P/m_{\tau}$, the inverse Lorentz transformation of Eq. (\ref{LTEnergy}) yields: 
\begin{equation}\label{x_approx2}
 x_{cms} \approx  \gamma x (1 + \beta \cos\theta).
\end{equation}
Using the condition $x=b(m_{\alpha})$ and substituting from Eq.~(\ref{x_approx2}) into Eq.~(\ref{xps}), we derive:
\begin{equation}\label{z_2body}
 x_{ps} = b(m_{\alpha})\left (1 + \beta \gamma^2 (1 + \beta u)v\right),
\end{equation}
where $u=\cos\theta$ and $v=\Delta \cos\theta$ capture the smearing due to the pseudo-rest frame boost.

The equation presented in Eq.~(\ref{z_2body}) reveals how the lepton energy in the pseudo-rest frame behaves as a function of the mass of the $\alpha$ particle. 
Two correlated random variables, $u$, and $v$, modulate this behavior. Within the pseudo-rest frame methodology, their distributions can be inferred from the particles on the tag side. Importantly, these distributions are independent of the signal side decay where we probe the new physics.

Assuming we can provide the joint PDF $f_{UV}(u,v)$, the marginal PDF $g(x_{ps};m_{\alpha})$ can be deduced by integrating out the inobservable variable $u$:
\begin{eqnarray}\label{gz_joint}
 g(x_{ps}; m_{\alpha}) = \int f_{UV}\left(u,\frac{x_{ps} - b(m_\alpha)}{\beta\gamma^2 b(m_\alpha)(1 + \beta u)}\right)\nonumber\\\times\frac{du}{|\beta\gamma^2 b(m_\alpha)(1 + \beta u)|}.
\end{eqnarray}
This is the normalized lepton energy distribution in the pseudo-rest frame as a function of the mass of the $\alpha$ boson.

\begin{figure}
\begin{center}
\includegraphics[width=8cm]{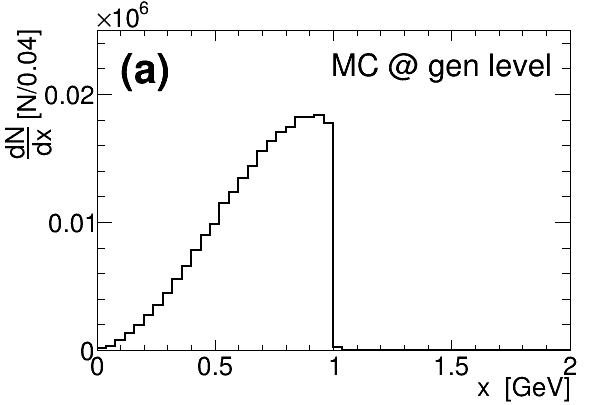}
\includegraphics[width=8cm]{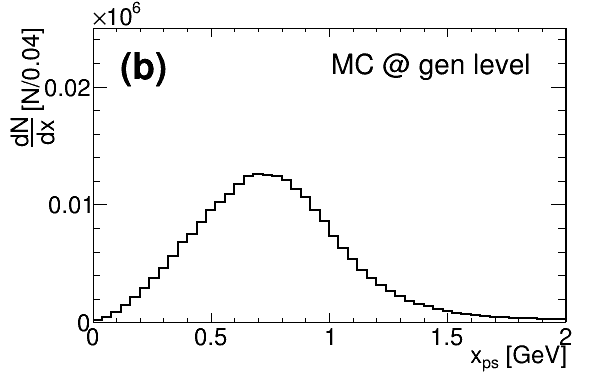}
\caption{Normalized lepton energy distribution for $\tau \to l \bar{\nu_{l}} \nu_{\tau}$ decays in the (a) tau rest frame and (b) the tau pseudo-rest frame.}\label{fig2} 
\end{center}
\end{figure}

For completeness, we also consider the three-body decay $\tau \to l + \bar{\nu_{l}}\nu_{\tau}$, where $l=e,\mu$. This decay, originating from the SM, serves as the primary background in the search for $\tau \to l+\alpha$.
Figure~\ref{fig2} visually portrays the $dN/dx$ distribution for the three-body decay, both in the tau rest frame (a) and in the pseudo-rest-frame (b), as obtained from Monte Carlo simulations at the generation level \footnote{For $\tau \to l + \bar{\nu_{l}}\nu_{\tau}$, $dN/dx
\sim x^2\left[3(1-x) + \frac{1}{2}(4x-3)\right]$.}.
Unlike the two-body decay case, the three-body decay does not require a $m_{\alpha}$ dependency. Therefore, from Eq. (\ref{DeltaX}), we choose $x_{ps}$ as the relevant random variable and $x$ as the variable to integrate due to its experimental inaccessibility. Then, the marginal PDF $g(x_{ps})$ is:
\begin{equation}\label{gz_joint_3body}
 g(x_{ps}) = \int f_{UV}(x,x_{ps} - x)dx.
\end{equation}
This is equivalent to Eq.~(\ref{gz_joint}) but tailored for the three-body decay case.

\section{Results}\label{sec:results}

We tested the accuracy of the PDFs given by Eq.~(\ref{gz_joint}) and Eq.~(\ref{gz_joint_3body}) through their implementation in the RooFit analysis software~\cite{Verkerke:2003ir}, part of the ROOT framework~\cite{Brun:1997pa}. The joint probability distributions $f_{UV}(u,v)$ were obtained from Monte Carlo simulations at the generation level of $\tau^{+}\tau^{-}\to (\pi^{+}\pi^{-}\pi^{+}\overline{\nu}_{\tau})(l + \bar{\nu_{l}}\nu_{\tau})$ produced in the open-source Belle II software~\cite{Kuhr:2018lps,basf2-zenodo}. We used two-dimensional histograms of variables $u=\cos\theta$ and $v=\Delta \cos\theta$. These histograms were subsequently smoothed with the k5b kernel algorithm available in ROOT. We then numerically calculated the integrals in Eq.~(\ref{gz_joint}) and Eq.~(\ref{gz_joint_3body}).

\begin{figure}[htb]
\centering
\includegraphics[width=8cm]{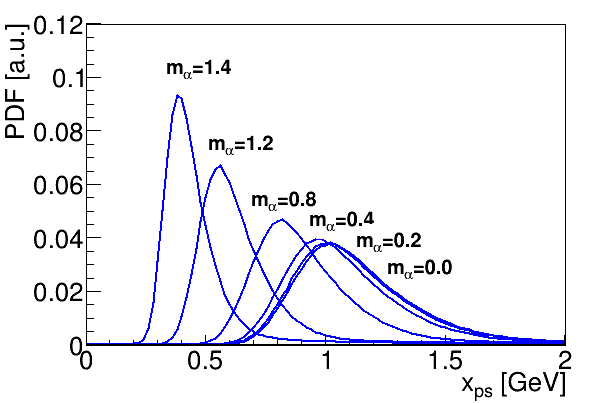}
\caption{PDF of normalized energy distribution in the pseudo rest frame for various $\alpha$ boson masses.}\label{gz_several_plots}
\end{figure}

Figure~\ref{gz_several_plots} displays the implemented function $g(x_{ps}; m_{\alpha})$ for several mass values of the $\alpha$ boson. As the mass approaches zero, the distributions become barely distinguishable and widen as $m_{\alpha}$ decreases.

\begin{figure}
\begin{center}
\includegraphics[width=8cm]{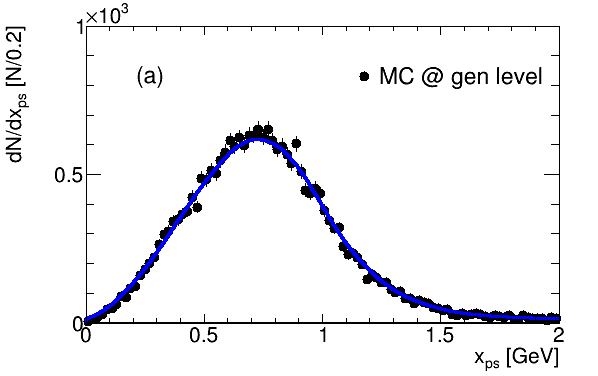}
\includegraphics[width=8cm]{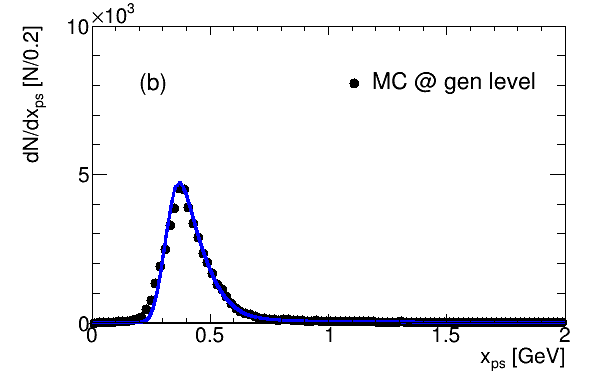}
\caption{PDF projection to the lepton energy distribution in the pseudo-rest frame from simulated data of (a) $\tau \to \ell \bar{\nu_{\ell}} \nu_{\tau}$, and (b) $\tau \to \ell+\alpha$ decays with $m_{\alpha}=1.4$ GeV. The solid line represents the implemented probability density functions in RooFit.}\label{dNdx_Fit_joint}
\end{center}
\end{figure}

Figure~\ref{dNdx_Fit_joint} illustrates that the implemented function accurately describes both (a) the three-body simulated data when boosted to the pseudo-rest-frame and (b) the example case of $\tau \to \ell+\alpha$ decays with $m_{\alpha}=1.4$ GeV.

\begin{figure}
\begin{center}
\includegraphics[width=8cm]{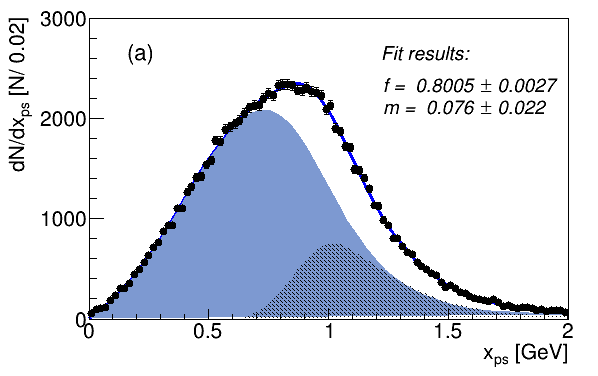}
\includegraphics[width=8cm]{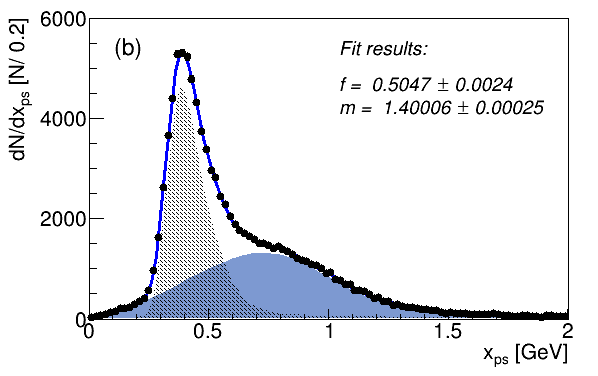}
\caption{Fitted parametric PDFs for $\tau\to l+invisible$ decays. Simulated data points are shown in black, comprising a mixture of $\tau \to \ell \bar{\nu_{\ell}} \nu_{\tau}$ and $\tau \to l+\alpha$ decay events. In example (a), the set parameters are a fraction $f=0.8$ of $\tau \to \ell \bar{\nu_{\ell}} \nu_{\tau}$ decays, and $\alpha$-boson mass $m=0.075$ GeV, while in example (b), they are $f=0.5$ and $m=1.4$ GeV. The solid blue area corresponds to SM events, the dash-dotted area to $\tau\to\ell+\alpha$ events, and the continuous line to their combined contribution.}\label{RooFitExample}
\end{center}
\end{figure}

In addition, to demonstrate how Eq.~(\ref{gz_joint}) and Eq.~(\ref{gz_joint_3body}) can be applied to the search for $\tau \to l+\alpha$ decays, Figure~\ref{RooFitExample} shows two examples of fits to simulated data. In both cases, the fit can accurately extract the mass of the $\alpha$ particle and its yield in the data.

\begin{figure}
\begin{center}
\includegraphics[width=8cm]{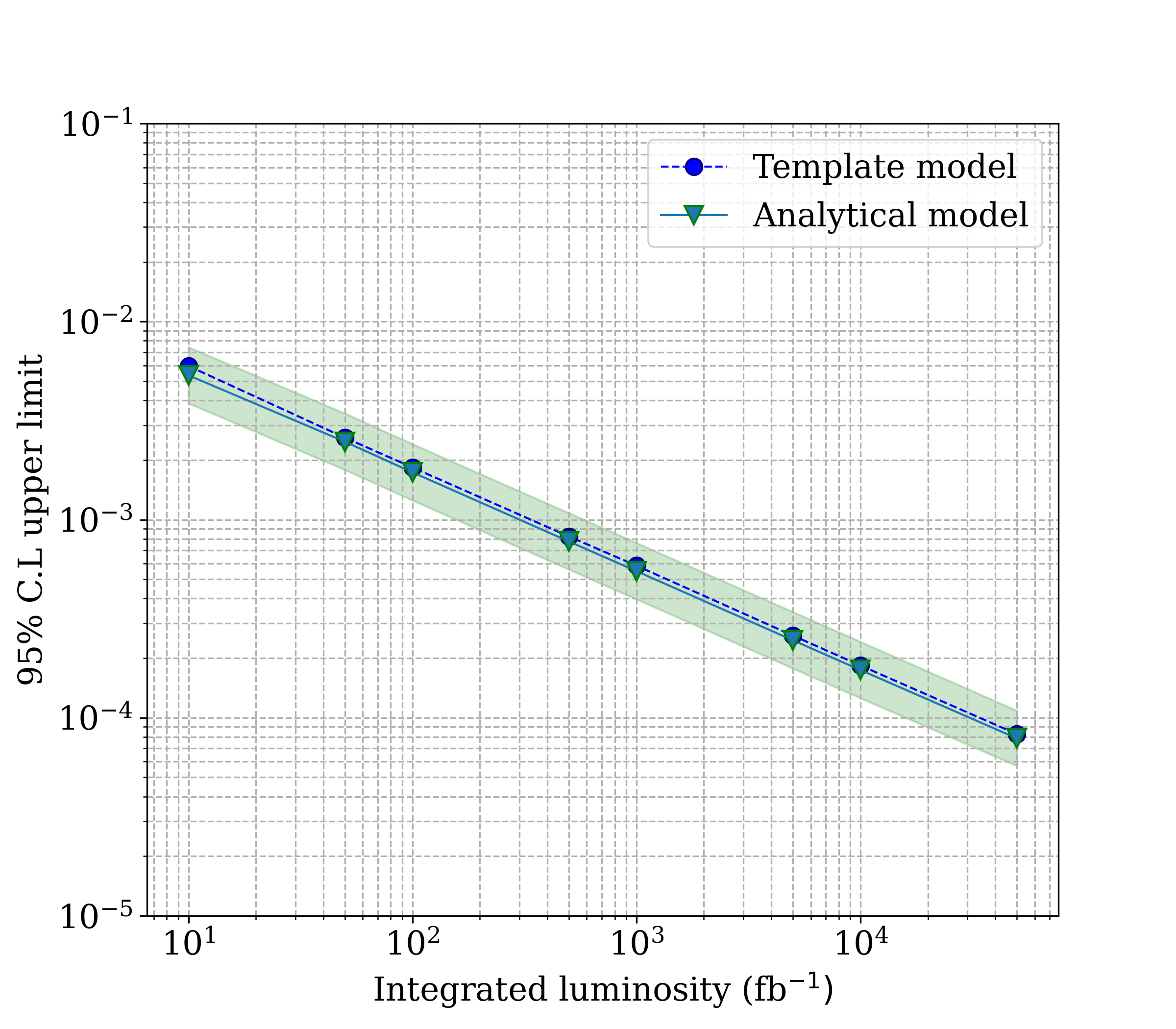}
\caption{95\% C.L. upper limits for the relative ratio of branching fractions \(B(\tau\to e\alpha)/B(\tau \to e \bar{\nu_{e}}\nu_{\tau})\) as a function of integrated luminosity for 3x1 prong tau pair decays. Blue circles (triangles) points are for the template (based on parametric PDFs) method. The green band represents one standard deviation for the parametric result. The upper limits were found for a mass of \(m_{\alpha} = 1.0\) GeV.}\label{UpperLimitsLum}
\end{center}
\end{figure}

Our approach further allows for replacing the histogram template method used by Belle II in its latest reported result with our functions. Using simulated data, we set an upper limit on the relative branching fraction for $m_{\alpha} = 1.0$ GeV and compared the outcomes. Figure~\ref{UpperLimitsLum} displays the results, showing both methods yield similar upper limits, with the advantage of not requiring specific mass value simulations with the implemented function.

\section{Conclusion\label{sec:conclusion}}

In this work, we have successfully developed a parametric probability density function that characterizes the LFV decay \(\tau \to \ell+\alpha\), as well as its primary irreducible background \(\tau \to l + \bar{\nu_{l}}\nu_{\tau}\). The specific formulations for these PDFs are presented in Eq.~(\ref{gz_joint}) and Eq.~(\ref{gz_joint_3body}).

We have demonstrated that the derived PDFs provide a reliable representation of the lepton energy distribution in the pseudo-rest frame. They overcome the limitations inherent in traditional template-based methods, facilitating a thorough examination of the full spectrum of allowable $\alpha$ boson mass values. Our findings indicate that these PDFs align with the template-based approach in searching for the $\alpha$ boson and yield comparable upper-limit estimations. This congruence enhances the reliability of our method and supports its applicability in addressing similar challenges in other particle searches.

A noteworthy observation, demonstrated in Figure~\ref{gz_several_plots}, is the distinct behavior of the distributions as the mass of the \(\alpha\) boson altered. As \(m_{\alpha}\) approached zero, the distributions converged to become virtually indiscernible, unveiling a critical aspect in the search for this particle.

The proposed PDF for the $\tau \to l + \alpha$ decay holds potential for further refinement via the input joint probability density function $f_{UV}$. For instance, incorporating observed physical backgrounds from decays like $\tau^{-}\to \pi^{-} \nu_{\tau}$, $\tau^{-}\to \pi^{-} \pi^{0}\nu_{\tau}$, and $\tau^{-}\to K^{-}\nu_{\tau}$ is feasible. Additionally, potential improvements can come from factoring in other smearing effects, such as detector resolution, or by integrating spin-dependent models for the $\alpha$ boson. Moreover, the potential development of a theoretical input for \(f_{UV}\) would represent a significant advancement despite the challenges associated with event-per-event correlations and the approximations guiding the pseudo-rest frame transformation.

Finally, the applicability of this work extends beyond the initial context of the two-body decay \(\tau \rightarrow l+\alpha\). For instance, it can be employed for exploring Dark Matter candidates or other processes involving a two-body decay with undetected particles, such as dark boson in the flavor-changing neutral current decay processes \(B^{\pm} \rightarrow h^{\pm}\chi\) \cite{Li:2021sqe}, where \(h\) stands for \(K^{+}, K^{0}_{s}, K^{\ast 0}, \pi^+, \pi^0 \rho^+ \) or \(\rho^0\), or to study a mechanism for low-temperature baryogenesis \cite{Elor:2018twp}, such as \(B^0 \rightarrow \Lambda^0 + \psi_{DM}\). 

This parametric function is a versatile tool suitable for any pseudo-rest frame method-related search, particularly when the decay involves undetected particles from the SM or beyond.

\section*{Acknowledgements}
We wish to thank I. Heredia de la Cruz for helpful discussions.
This work was supported by the Conahcyt research grant CB 320328.

\bibliographystyle{elsarticle-num} 
\bibliography{references.bib}

\end{document}